\documentstyle[epsf,aps,pre]{revtex}

\newcommand{\be}{\begin{equation}}
\newcommand{\ee}{\end{equation}}
\newcommand{\bea}{\begin{eqnarray}}
\newcommand{\eea}{\end{eqnarray}}

\begin{document}

\bibliographystyle{/usr/local/lib/tex/macros/prsty}

\twocolumn[\hsize\textwidth\columnwidth\hsize\csname
@twocolumnfalse\endcsname

\title{Defect Chaos of Oscillating Hexagons in Rotating Convection}

\author{Blas Echebarria and Hermann Riecke}

\address{
Department of Engineering Sciences and Applied Mathematics,\\
Northwestern University, 2145 Sheridan Rd, Evanston, IL, 60208, USA
}

\date{\today}

\maketitle

\begin{abstract}
Using coupled Ginzburg-Landau equations, the dynamics of hexagonal patterns 
with broken chiral symmetry are investigated, as they appear in rotating 
non-Boussinesq or surface-tension-driven convection. We find that close to the 
secondary Hopf bifurcation to oscillating hexagons the dynamics are well described by a single complex Ginzburg-Landau equation (CGLE) coupled to the phases of the hexagonal pattern. At the bandcenter these equations reduce to the usual CGLE
and the system exhibits defect chaos. Away from the bandcenter a transition to a frozen vortex state is found.
\end{abstract}

]


The interplay between theory and experiments is at the heart of the progress
made in recent years in the study of pattern-forming systems far from equilibrium. Thus,
investigations of specific physical systems that allow a quantitative comparison
with precise experiments are essential to check available
theories and often also to suggest new theoretical venues. This has been particularly
successful for ordered patterns (e.g. in
Rayleigh-B\'enard convection and Taylor-Couette flow \cite{Ah92}). For disordered patterns
and spatio-temporal chaos a number of different systems have been under investigation
with the goal to achieve detailed comparison between experiments and theoretical work.
Spiral-defect chaos
in Rayleigh-B\'enard convection with low Prandtl number exhibits very rich dynamics
(e.g. \cite{MoBo96}).
However, they do not arise at small amplitudes and are therefore not accessible by
weakly nonlinear theory. In the presence of rotation the K\"uppers-Lortz instability
of convection rolls induces another type of spatio-temporal chaos (e.g. \cite{HuPe98}).
Although it occurs directly at onset a systematic theoretical treatment
within weakly nonlinear theory has not
been successful due to the isotropy of the system.
In anisotropic electroconvection of nematic liquid crystals spatio-temporal
chaos \cite{DeAh96} can be systematically described by coupled Ginzburg-Landau equations.
 The derivation of these equations from the microscopic equations is, however,
extremely involved and not yet complete \cite{TrKr97}.

From a theoretical point of view the most attractive and therefore
most extensively studied canonical equation exhibiting spatio-temporal chaos
 is the complex Ginzburg-Landau equation (CGLE). It describes
the onset of oscillations in a spatially extended system.
Despite its simplicity it exhibits an extraordinary
variety of complex dynamics, including phase chaos, defect chaos, and
frozen structures \cite{ChMa96,MaCh96,GrJa96,MoHe96,MoHe97,To96,ToFr97,PoSt95,HuAl92}.
In the one-dimensional case many aspects of its dynamics have been observed
 experimentally (e.g. localized solutions \cite{BuCh99}).
In two dimensions, however, no detailed comparison of any of the theoretical
regimes of complex dynamics with experiments is available.

  In this Letter we suggest that weakly nonlinear hexagon patterns in
rotating convection are a good candidate to compare the theoretical results for
spatio-temporal chaos in the two-dimensional CGLE with experiments.
Due to the rotation the hexagons typically undergo a
transition to oscillating hexagons, which are described by a complex Ginzburg-Landau
equation coupled to two phase equations. We show that in a sufficiently large system the
oscillating hexagons exhibit a state of defect chaos that is well described by the
usual two-dimensional CGLE. Depending on the wavenumber of the
underlying hexagons we find a transition to a frozen vortex state, as it is also
the case in the CGLE.

We consider three coupled amplitude equations, describing the dynamics 
of hexagonal patterns with broken chiral symmetry, as they appear in rotating 
non-Boussinesq or surface-tension-driven convection. These equations can be 
obtained from the corresponding physical equations (e.g. 
Navier-Stokes) for small amplitudes with the usual scalings \(  \partial _{t}\sim {\mathcal{O}}(A_i^2) \) and \( \nabla 
\sim {\mathcal{O}}(A_i) \), provided that the coefficient of the quadratic resonant term is small. After rescaling they can be written as:
\begin{eqnarray}
\partial _{t}A_{1} & = & \mu A_{1}+({\bf {n}}_{1}\cdot \nabla )^{2}A_{1}+A^{*}_{2}A^{*}_{3} -  A_{1}|A_{1}|^{2} \label{eq.amp} \\
&&-(\nu +\gamma )A_{1}|A_{2}|^{2}-(\nu -\gamma )A_{1}|A_{3}|^{2}.\nonumber  
\end{eqnarray}
The equations for the other two amplitudes are obtained by 
cyclic permutation of the indices. The broken chiral 
symmetry  manifests itself by the cross-coupling coefficients not being
equal. The coefficient $\gamma$ is, therefore, a measure of the rotation.

A study of the steady hexagons with a wavenumber ${\bf k}$ slightly different from the critical wavenumber (${\bf k}_n={\bf k}_n^c + {\bf q}_n$, $|{\bf q}_n|=q$) and their sideband instabilities has been undertaken in \cite{EcRipre}. We focus here on the secondary Hopf bifurcation that appears for larger values of the control parameter $\mu=\mu_c \equiv (2+\nu)/(\nu-1)^2 + q^2$ \cite{Sw84,So85}. It gives rise to what we call oscillating hexagons, in which the three amplitudes of the hexagonal pattern oscillate with frequency \( \omega=2\sqrt{3}\gamma/(\nu -1)^{2} \) and a phase shift of $2\pi /3$ among them.   As  \( \mu \) is increased further, eventually a 
point $\mu=\mu_{het}$ is reached at which the branch of oscillating hexagons ends on the branch corresponding to an unstable mixed-mode solution in a global bifurcation involving a heteroclinic connection \cite{Sw84,So85,MiPe92}. Above this point the only stable 
solution is the roll solution. When $|\gamma| > \nu-1$ the rolls are never stable and the 
limit cycle persists for arbitrarily large values of $\mu$. In the absence 
of the quadratic term in Eq. (\ref{eq.amp}) this condition corresponds to the 
K\"uppers-Lortz instability of rolls \cite{KuLo69}. Far above the Hopf bifurcation 
the periodic orbit is expected to become anharmonic, and may somewhat resamble 
the state encountered in the K\"uppers-Lortz regime of rotating 
Boussinesq Rayleigh-B\'enard convection \cite{BuCl79a,BuHe80}.

To study the stability of the oscillating hexagons close to the Hopf bifurcation point, the amplitudes $A_n$ are expanded as:
\begin{equation}
A_n=(R+[ e^{2\pi ni/3}\sqrt{\epsilon}{\cal H}e^{i\omega t} + c.c.] + {\cal O}(\epsilon))e^{i{\bf q}_n \cdot {\bf x} +\sqrt{\epsilon}\phi_n}. 
\end{equation}
The amplitude of the steady hexagons at the bifurcation point, $R=1/(\nu-1)$, is independent of $\mu$ and $q$. Since we are considering a secondary bifurcation the amplitude ${\cal H}$ will couple to the phases of the hexagonal pattern \cite{CoIo90}. From the phases $\phi_n$ of each of the modes it is possible to construct a phase vector $\vec{\phi}=(\phi_x, \phi_y)\equiv (-\phi_2-\phi_3,(\phi_2-\phi_3)/\sqrt{3})$, whose components are related 
to translations in the x- and y-directions \cite{LaMe93,Ho95,EcPe98}. Eliminating the fast variables, we
arrive at order $\epsilon^{3/2}$ at an equation for the amplitude of the oscillation ${\cal H}$, coupled to the
phases of the underlying hexagonal pattern,
\begin{eqnarray}
\partial_T{\cal H}&=&\varepsilon\delta_1{\cal H}+\xi\nabla^2{\cal H}-\delta_2{\cal H}
\nabla\cdot\vec{\phi}-\rho{\cal H}|{\cal H}|^2, \label{eq.cgle-ph.a}\\
\partial_T \vec{\phi}&=&D_{\bot}\nabla^2 \vec{\phi}+D_{\|}\nabla(\nabla\cdot
\vec{\phi})+D_{\times_1}(\hat{\bf e}_z\times\nabla^2\vec{\phi}) \label{eq.cgle-ph.b}
\\
&&+D_{\times_2}(\hat{\bf e}_z\times\nabla)(\nabla\cdot\vec{\phi})+\alpha\nabla|{\cal H}|^2\nonumber \\
&&+\beta_1(\hat{\bf e}_z\times\nabla)|{\cal H}|^2
-i\beta_2({\cal H}\nabla {\cal H}^{*} - {\cal H}^{*}\nabla{\cal H})\nonumber \\
&&+i\eta[{\cal H}(\hat{\bf e}_z\times\nabla)
{\cal H}^{*}-{\cal H}^{*}(\hat{\bf e}_z\times\nabla){\cal H}], \nonumber
\end{eqnarray}
where $\hat{\bf e}_z$ is the unit vector in the direction perpendicular to the plane, $\varepsilon=\mu-\mu_c$, $\omega=2\sqrt{3}\gamma R^2$ and
\begin{eqnarray*}
&&v=3R(1+2R),\\
&&\delta_1=\frac{2R}{v}-\frac{2i\omega}{v},\;\;\delta_2=q\delta_1,\\
&&\xi=\frac{1}{2} - \frac{3q^2 R}{9R^2 + \omega^2} - \frac{iq^2}{\omega}\frac{9R^2+2\omega^2}{9R^2 + \omega^2},\\ 
&&\rho=\frac{8(3R+1)}{v}-\frac{4i\omega (1+4R)}{Rv} -\frac{32i}{3\omega},\\
&&D_{\bot}=\frac{1}{4},\;\;D_{\|}=\frac{1}{2} -\frac{2q^2}{v},\;\;D_{\times_1}=\frac{q^2}{\omega},\;\;D_{\times_2}=0,\\
&&\alpha=-\frac{2q\omega^2}{9R^2 + \omega^2} - \frac{2q(1+6R)}{Rv},\\
&&\beta_1=\frac{6\omega q}{R(9R^2+\omega^2)},\;\;\beta_2=\beta_1,\;\;\eta=\frac{18q}{9R^2+\omega^2}.
\end{eqnarray*} 
It is worth pointing out that the phase-amplitude equations 
(\ref{eq.cgle-ph.a},\ref{eq.cgle-ph.b}) can be deduced by means of
symmetry arguments alone and are, therefore, generic to this order in 
$\epsilon$. In fact, they could be derived directly from the fluid equations without the use of the Ginzburg-Landau equations (\ref{eq.amp}). Thus, keeping higher order terms in  (\ref{eq.amp})  would change the values of the coefficients in (\ref{eq.cgle-ph.a},\ref{eq.cgle-ph.b}), but not their form.  

\begin{figure}
\centerline{\epsfxsize=8cm \epsfbox{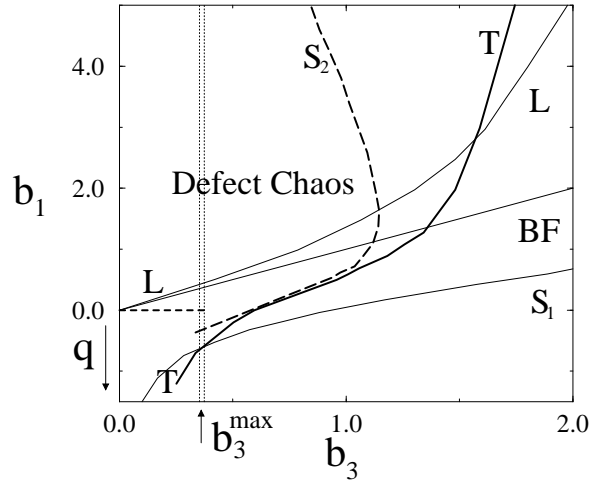}} 
\caption{Phase diagram for the CGLE (after [6]). Defect chaos is found to the left of line T. Above the Benjamin-Fair (BF) line, all plane waves are longwave unstable, resulting in a state of phase chaos, which is bistable with defect chaos or a frozen vortex state up to the line L. The lines $S_1$ and $S_2$ represent the convective and absolute stability limits of plane waves emitted by spirals. The vertical dotted lines represent the limits for the range of $b_3^{\rm max}$ (cf. (\ref{eq.b3})). As $q$ is increased $b_1$
decreases.}
\label{fig.diag}
\end{figure}

Central to the results presented in this Letter is the observation that for hexagons with
rotation as described by (\ref{eq.amp}) the phase-amplitude equations (\ref{eq.cgle-ph.a},\ref{eq.cgle-ph.b})
 decouple at the bandcenter, i.e. for $q=0$. In this 
case they reduce to the usual CGLE for the amplitude of the oscillation. 
Scaling the 
time, space and amplitude, Eq. (\ref{eq.cgle-ph.a}) can be written in the more usual form \cite{ChMa96}:
\begin{equation}
\partial_t H = H + (1+ib_1) \nabla^2 H - (b_3 -i \,{\rm sign}(\omega)) H |H|^2
\label{eq.cgle}
\end{equation}
where $b_1=\xi_i/\xi_r=0$ and $b_3$ is given by:
\begin{equation}
b_3=\frac{\rho_r}{|\rho_i|}=\frac{2|\omega| R (3R+1)}{\omega^2 (1+4R) + 8R^2 (1+2R)}.
\label{eq.b3}
\end{equation}
As $\omega$ is increased by increasing the rotation rate $\gamma$, $b_3$ reaches a maximum $b^{\rm{max}}_3=b^{\rm{max}}_3(\nu)$ at
$\gamma^{\rm{max}}=\sqrt{2(\nu-1)^2(\nu+1)/3/(\nu+3)}$. For any value of $\nu$ the coefficient $b_3$ will move across
the range $0 < b_3 < b^{\rm{max}}_3$ as the rotation rate is varied. Note that the dependence of $b^{\rm{max}}_3$ on $\nu$ is only weak and is limited to
the range $0.354 < b^{\rm{max}}_3 < 0.375$, as indicated by the dotted lines in
Fig. \ref{fig.diag}. Thus, independent of  $\nu$, which depends on the physical properties of the system, the
oscillating hexagons are always
in an intermittent regime in which stable plane waves and defect chaos coexist (see Fig. \ref{fig.diag}).

It should be noted that, although we are
considering a system with broken chiral symmetry, to this order in the perturbation expansion we obtain the usual CGLE, which is chirally symmetric.
Going to higher order we would expect terms like $\hat{\bf e}_z \cdot (\nabla H \times \nabla H^{*}) H$ to appear in Eq. (\ref{eq.cgle}), breaking the 
symmetry between spirals with opposite topological charge \cite{NaOt98}.

\begin{figure}
\centerline{\epsfxsize=4cm\epsfbox{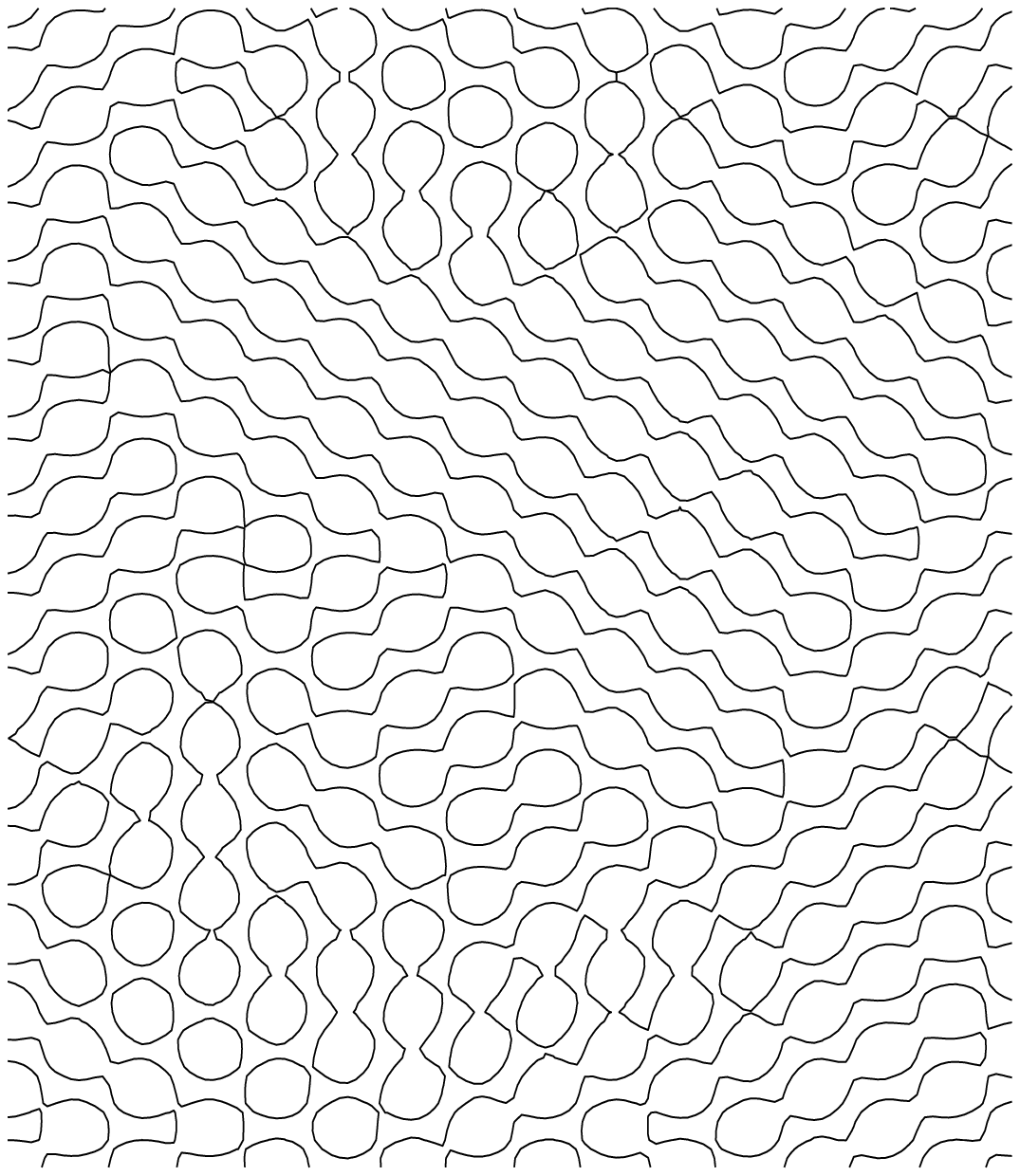}\hspace{0.5cm}\epsfxsize=4cm\epsfbox{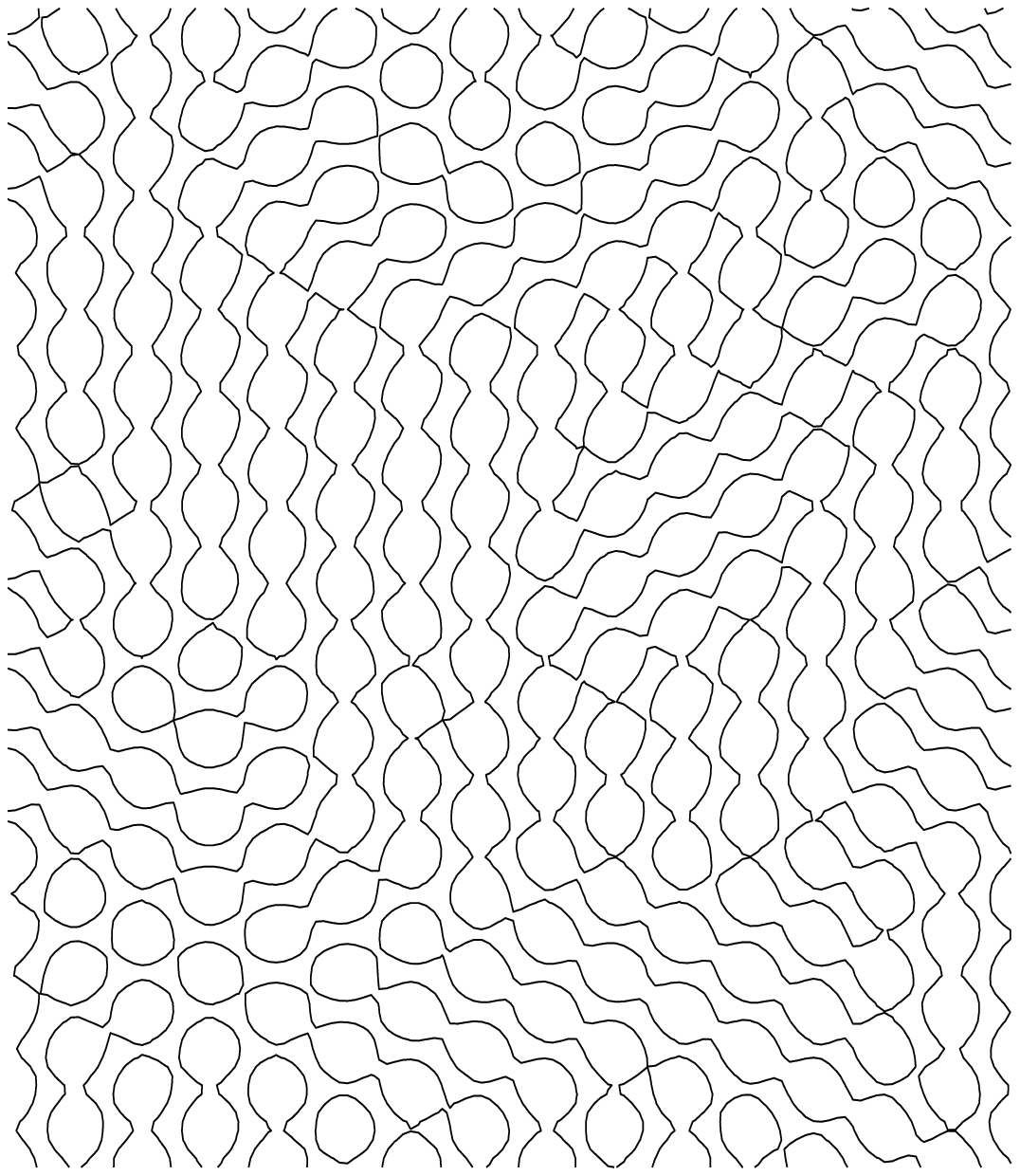}}
\vspace{0.5cm}
\caption{Reconstruction of the hexagon pattern $\psi=\sum^{3}_{j=1} A_j e^{i{\bf k}^c_j \cdot {\bf x}}$, with $k_c=30\pi/L_x$, obtained by simulating numerically Eqs. (\ref{eq.amp}) in a box of length $L_x=50$, $L_y=100/\sqrt{3}$ with
$128\times 128$ modes, for $\mu=4.6$, $\nu=2$,
$\gamma=0.5$ and $q=0$. The contour lines are taken at $\psi=-0.7$. The time    difference between the two snapshots is half a period of oscillation.
\label{fig.hexa}}
\end{figure}

\begin{figure}
\centerline{\epsfxsize=4cm \epsfbox{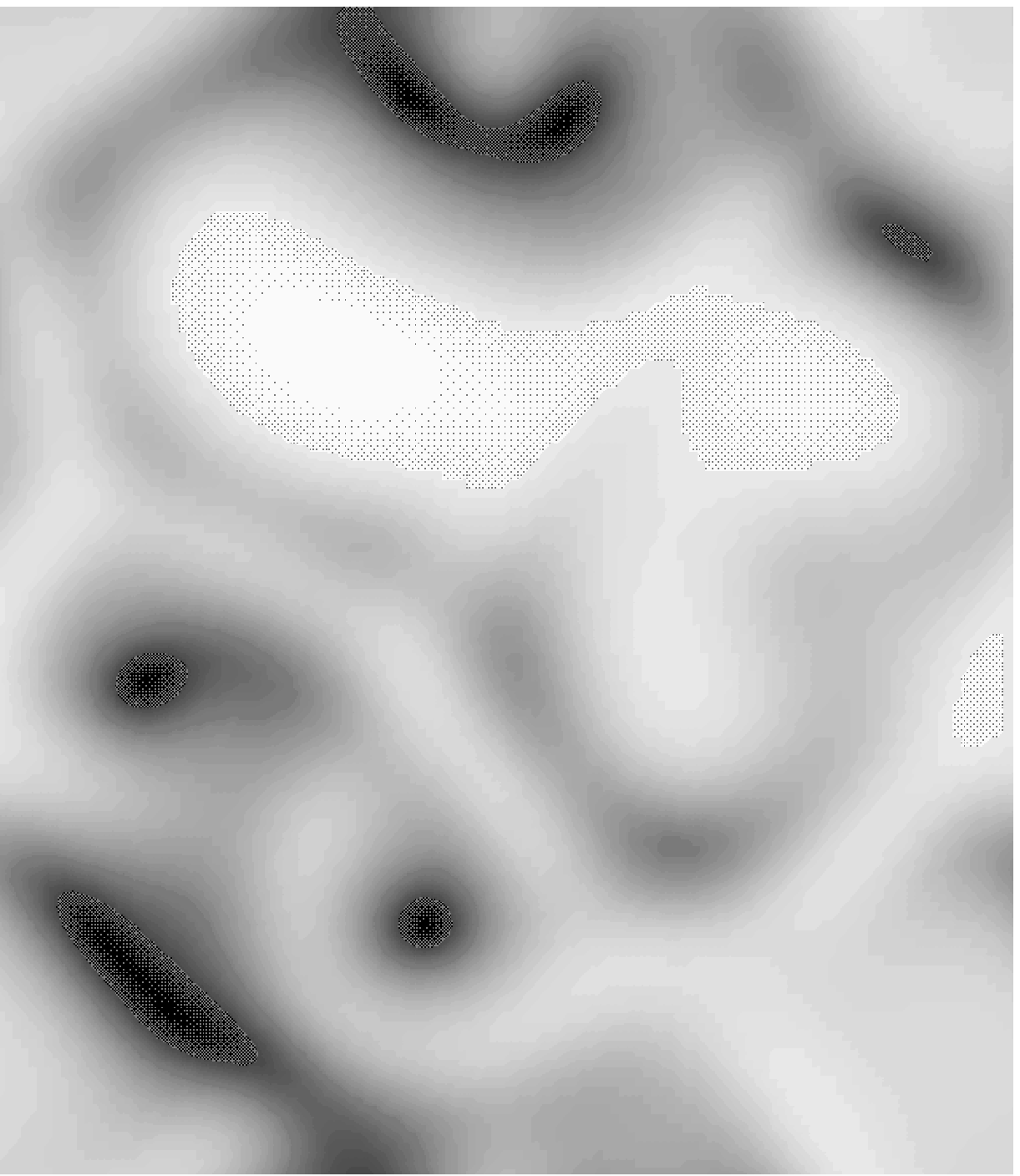}\hspace{0.5cm}\epsfxsize=4cm \epsfbox{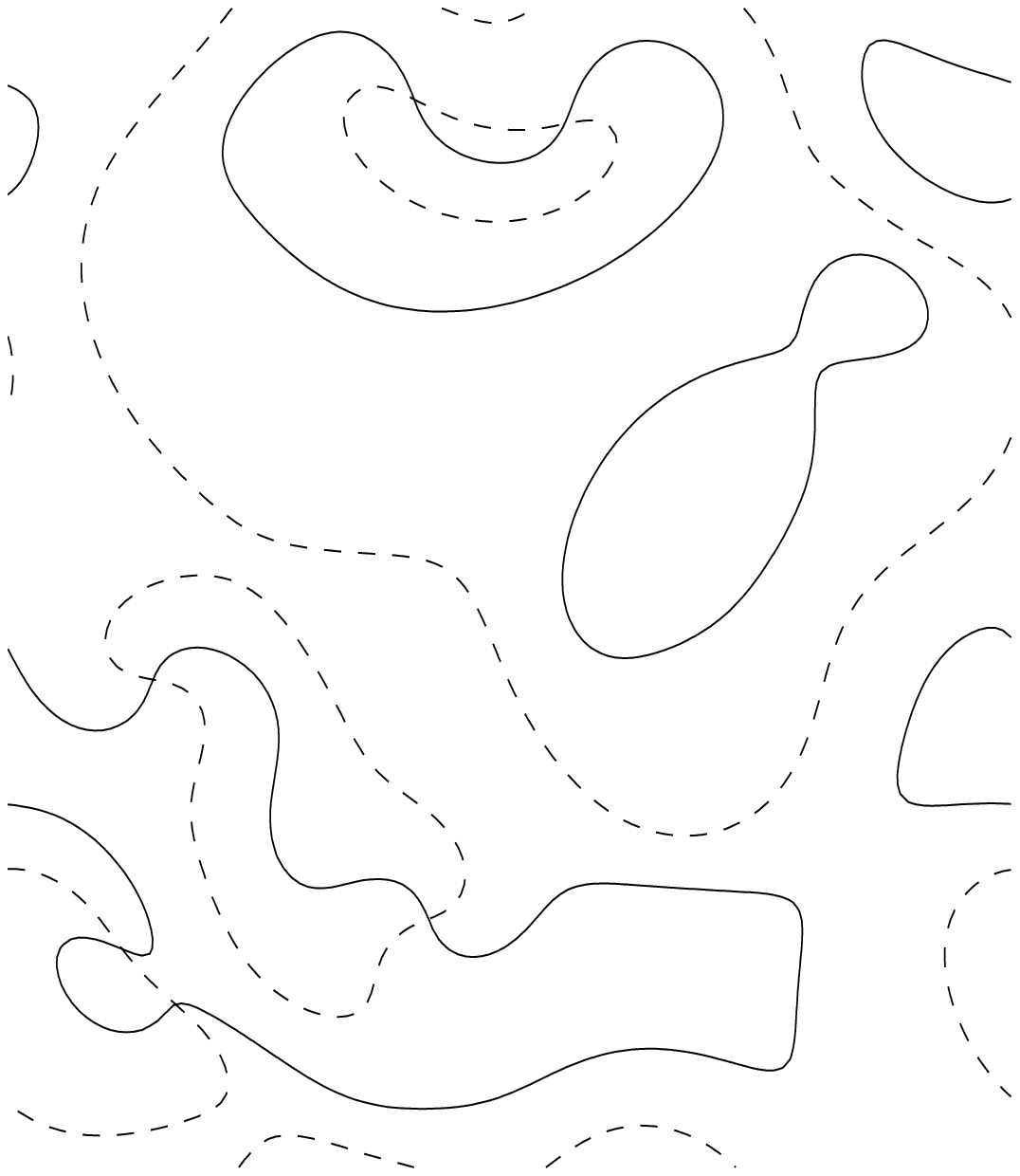}}
\caption{Snapshots corresponding to Fig. \ref{fig.hexa} of the a) modulus of ${\cal H}$, b) lines ${\rm Re}({\cal H})=0$ (solid) and ${\rm Im}({\cal H})=0$ (dashed). \label{fig.snap}}
\end{figure}

We investigate the defect chaos regime by numerical simulations of Eq. (\ref{eq.amp}) using a pseudospectral method with a 4th-order Runge-Kutta/integrating factor time-stepping scheme and periodic boundary conditions. To allow for regular hexagonal patterns we take a rectangular box 
of aspect ratio $L_x/L_y = \sqrt{3}/2$. Fig. \ref{fig.hexa} shows a picture of the hexagonal pattern in this regime. It consists of patches of slightly roll-like hexagons, whose preferred direction oscillates on a fast time scale. The patches change shape and size on a slow time scale. The
regions where almost perfect hexagons can be observed (e.g. on the bottom left part of the figure) correspond to the zeros (defects) of the oscillation amplitude.
In order to extract this complex amplitude ${\cal H}$ from the amplitudes $A_n$ we use that close to onset ${\cal H}e^{i\omega t} + {\cal H}^{*}e^{-i\omega t}
\simeq |A_n|-(\sum^3_{j=1} |A_j|)/3$. The amplitude ${\cal H}$ is obtained 
multiplying the former expression by 
$e^{-i\omega t}$ and taking the average over each period. A snapshot of the magnitude $|{\cal H}|$ as obtained from Fig. \ref{fig.hexa} is given in Fig. \ref{fig.snap}a, while in Fig. \ref{fig.snap}b we show the corresponding lines ${\rm Re}({\cal H})=0$ and ${\rm Im}({\cal H})=0$, whose intersection points correspond to the
defects of ${\cal H}$.

Away from the bandcenter, $q\neq 0$, $b_1$ becomes non-zero and is given by:
\begin{equation}
b_1=\frac{2(9R^2+2\omega^2)q^2}{(6q^2R-9R^2-\omega^2)\omega}.
\end{equation}
The coefficient $b_3$ remains unchanged.
Depending on the value of $q$ the system can therefore cross the line T  (see Fig. \ref{fig.diag}), beyond which defect chaos is no longer an attractor within the CGLE (\ref{eq.cgle}). But, for $q\neq 0$, the oscillation amplitude ${\cal H}$ is coupled to the phase $\vec{\phi}$. In order to study the influence of this coupling in the defect chaotic regime we measure the density of defects for Eqs. (\ref{eq.amp}) and Eqs. (\ref{eq.cgle-ph.a},\ref{eq.cgle-ph.b}) as a function
of $q$ (accordingly $b_1$) and compare it with the results for the CGLE (\ref{eq.cgle}).

We begin with a perfect steady hexagonal pattern with wavenumber $q$ ($A_n=Re^{i{\bf q}_n\cdot {\bf x}}$, $|{\bf q}_n|=q$, in Eq. (\ref{eq.amp}), 
${\cal H}=0$, $\vec{\phi}=0$ in Eqs. (\ref{eq.cgle-ph.a},\ref{eq.cgle-ph.b}), 
$H=0$ in Eq. (\ref{eq.cgle})) and add noise of zero mean. If the system is large enough the resulting oscillating state is not homogeneous but ends up in a 
persistent chaotic state. We also checked numerically that a perfect oscillating state in Eq. (\ref{eq.amp}), $A_n=(R+{\cal H}e^{i\omega t}+c.c.)e^{i{\bf q}_n\cdot {\bf x}}$, is linearly stable, although sufficiently strong perturbations can destabilize it, giving rise to defect chaos. In the
 simulations we use $64\times 64$ modes, in a box of length $L_x=200$ for Eqs. (\ref{eq.amp}) and (\ref{eq.cgle-ph.a},\ref{eq.cgle-ph.b}), and $L=50$ for the CGLE, which roughly
corresponds to the former after rescaling. As the scaling depends on $q$ we 
take the same box size $L_x , L_y$ in all the simulations of Eqs. (\ref{eq.amp}) and (\ref{eq.cgle-ph.a},\ref{eq.cgle-ph.b}) and then scale the density of defects appropriately to compare with the results from the CGLE.   
After a transient time, the number of defects is measured for a sufficiently long time so the system is statistically stationary. The results
corresponding to the CGLE are the average of three independent runs. 

\begin{figure}
\centerline{\epsfxsize=8cm \epsfbox{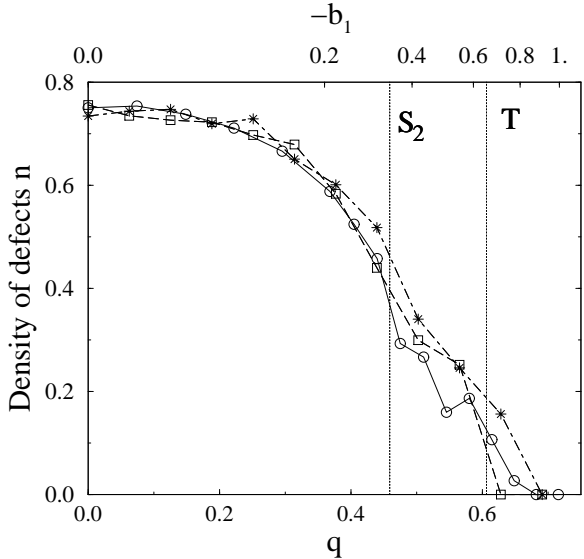}} 
\caption{Density of defects $n$ as a function of the wavenumber of the underlying pattern $q$ for $b_3=0.355$ ($\nu=2$, $\gamma=0.5$, $\mu=\mu_c + 0.1$). The circles correspond to the simulations of Eq. (\ref{eq.cgle}), while the squares have been obtained simulating Eqs. (\ref{eq.cgle-ph.a},\ref{eq.cgle-ph.b}). The stars are the results obtained 
with the amplitude equations (\ref{eq.amp}). Between lines $S_2$ and $T$ the system ends up in a frozen state.
\label{fig.def}}
\end{figure}

As can be seen in Fig. \ref{fig.def} the simulations of (\ref{eq.amp}), (\ref{eq.cgle-ph.a},\ref{eq.cgle-ph.b}) and (\ref{eq.cgle}) agree very well up to the line $S_2$, where the spiral defects go from being absolutely to convectively unstable. Since for $q\neq 0$ the amplitude ${\cal H}$ is no longer decoupled from the phases, the agreement indicates that the phase does not 
have a strong influence on the dynamics in this regime. Beyond line $S_2$, after a chaotic transient, the system settles down in a frozen vortex state  in which the density of defects remains unchanged for long periods of time. Since this final state is history dependent a quantitative agreement is harder to achieve. In particular it would require a more extensive sampling of initial conditions. However, the agreement remains good. Beyond line T still a small number of defects remain, probably because our simulations are not long enough given the weak interaction of defects. If and to what extent the coupling with the phase affects the lines $S_2$ and $T$ is not clear from these simulations, but the qualitative picture remains 
the same as in the CGLE. Note also that, for these parameters, additional long-wave instabilities occur beyond line $T$ \cite{EcRiunpub}.

In conclusion, we have shown that within weakly nonlinear theory
hexagons arising in rotating convection are a good candidate to investigate
two-dimensional defect chaos. Due to the rotation the hexagons undergo a
Hopf bifurcation to oscillating hexagons. Although these are usually linearly stable
at the band center, finite perturbations that excite defects lead to persistent
spatio-temporal chaos. In this regime the single complex Ginzburg-Landau
equation describes the dynamics quantitatively. Further away from the band center we find a transition to a
frozen vortex state, which is also in agreement with the CGLE. The decoupling of the phases
and the amplitude that occurs at the band center will be modified if higher-order
corrections are taken into account in the amplitude equations (\ref{eq.amp}).
Simulations in which nonlinear gradient terms in (\ref{eq.amp})
are included suggest, however, that the qualitative picture
remains the same \cite{EcRiunpub}.
It is worth noting that the dynamics discussed in this paper (in particular, the spatio-temporally chaotic state) have been obtained for values of the rotation rate below the K\"uppers-Lortz
instability. Therefore we expect that this genuinely new regime of rotating 
convection is accessible in currently available experiments. We hope that these results
will trigger new experiments, which will contribute to a better understanding
of the mechanisms of transition to spatio-temporal chaos.

We gratefully acknowledge interesting discussions with F. Sain and M. Silber.
The numerical simulations were performed with a modification of a code by G.D. Granzow.
This work was supported by the Engineering Research Program of the
Office of Basic Energy Sciences at the Department of Energy (DE-FG02-92ER14303)
 and NASA Grant NAG3-2113.



\end{document}